\documentclass[copyright,creativecommons]{eptcs}
\providecommand{\event}{DCM 2012}
\usepackage{breakurl}             % Not needed if you use pdflatex only.

\usepackage{amsmath,amssymb,amsthm}
\usepackage{enumerate}
\usepackage{algorithm}
\usepackage{textcomp}
\usepackage{color}
\usepackage{stmaryrd,bussproofs}
\usepackage{dsfont,booktabs}
\usepackage{helvet}

\renewcommand{\emptyset}{\varnothing}

\newcommand{\STOP}{\mathrm{STOP}}
\newcommand{\SKIP}{\mathrm{SKIP}}
\newcommand{\THROW}{\mathrm{THROW}}
\newcommand{\YIELD}{\mathrm{YIELD}}
\newcommand{\SKIPP}{\mathrm{SKIPP}}
\newcommand{\THROWW}{\mathrm{THROWW}}
\newcommand{\YIELDD}{\mathrm{YIELDD}}

\title{General dynamic recovery for compensating CSP}
\author{Abeer S. Al-Humaimeedy
\institute{Department of Informatics\\King's College London, United Kingdom}
\institute{Department of Information Technology\\King Saud University, Riyadh, Saudi Arabia}
\email{Abeer@ksu.edu.sa}
\and
Maribel Fern\'andez
\institute{Department of Informatics\\
King's College London, United Kingdom}
\email{Maribel.Fernandez@kcl.ac.uk}
}
\def\titlerunning{General dynamic recovery for compensating CSP}
\def\authorrunning{A.\ S.\ Al-Humaimeedy \& M.\ Fern\'andez }
\begin{document}
\maketitle

%--------------------------------------------------------------------ABSTRACT-START
\begin{abstract}
Compensation is a technique to roll-back a system to a consistent
state in case of failure.  Recovery mechanisms for compensating
calculi specify the order of execution of compensation
sequences. Dynamic recovery means that the order of execution is
determined at runtime.
In this paper, we define an extension of Compensating CSP, called
DEcCSP, with general dynamic recovery. We provide a formal,
operational semantics for the calculus, and illustrate its expressive
power with a case study.  In contrast with previous versions of
Compensating CSP, DEcCSP provides mechanisms to replace or discard
compensations at runtime.  Additionally, we bring back to DEcCSP
standard CSP operators that are not available in other compensating
CSP calculi, and introduce channel communication.

\end{abstract}
%--------------------------------------------------------------------ABSTRACT-END

%----------------------------------------------------------------------DOCUMENT-START
\section{Introduction}

Transactions are units of work comprising a set of interactions between entities to achieve a final output~\cite{tbook}. The notion of a transaction is based on the idea of all-or-nothing, that is, none of the transaction's effects can take place until the whole transaction is committed.

Basically, there are two types of transactions: \emph{Atomic Transaction (AT)} and \emph{Long Running Transaction (LRT)}~\cite{tbook}. ATs prevent entities from updating system resources until the whole transaction is committed. Checkpoints and resources' key-locks are used to maintain systems in a safe state. LRTs relax the previous condition and allow the entities to update resources. However, LRTs use compensations to maintain systems in a safe state.  Compensation is a technique to roll-back the system to a consistent state in the case of failure.

LRTs are intensively used in complex systems where entities usually engage in transactions that last for hours, days or even longer while resources cannot be locked for such a long time. As a result, modelling languages for complex systems should be equipped with techniques to implement LRTs.

Considering process calculi as modelling languages, compensation has emerged as a crucial need. The fundamental idea behind compensating process calculi is to adapt the well-developed transaction techniques from database theory to the theory of process calculi, by introducing primitives to model and handle transactions. In essence, the key concepts introduced within process calculi are: scope, fault, termination and compensation. \textit{Scope} defines the transaction boundaries. Issues to be considered in relation with scope are: the relation between transaction's subprocesses and the fate of subtransactions if the parent transaction terminates (if nested transactions are allowed). Subtransactions can be aborted (i.e., terminated), discarded (i.e., deleted), or preserved (i.e., subtransactions are levelled up and continue running).
\textit{Fault} represents an exception (internal fault) thrown by a process; fault handlers are procedures that should be evaluated in such a case. \textit{Termination} is the state of a process which is either committed or interrupted by other processes (external fault); termination handlers are procedures that should be evaluated in such a case.
Finally, \textit{compensation} is the reverse behaviour of a process, to undo the effects of the normal execution in the case of a failure. Compensations are evaluated when the process needs to roll-back. Issues to be considered when dealing with compensation are the installation of a new compensation in the system, and the recovery mechanism which determines the evaluation order of the compensations to recover the system to a consistent state after a failure.

Basically, compensation can be statically implemented in any process calculus by creating a new process, which captures a fixed compensation scenario (planned in advance). Static compensation is feasible in systems where the evaluation of processes is fixed. However, in complex systems where there are interleaved, parallel and complex patterns of interactions, the compensation scenario heavily depends on the execution order. Therefore, compensating process calculi have been proposed as a suitable solution for modelling such complex systems. The fundamental idea of such process calculi is introducing a new type of processes called \emph{compensable processes}.

A compensable process comprises two behaviours:  the \emph{forward behaviour} corresponds to the normal execution of the process, and the \emph{compensation behaviour} corresponds to its reverse execution, which will undo the effects of the normal execution in case of system failure. While the system is running, a sorted compensation scenario (sorted according to the execution order) will be built incrementally from individual reverse behaviours.

Compensation has been introduced in a range of process calculi, including CCS~\cite{cccs}, $\pi$-calculus~\cite{webpi,dcpi,l2011}, CSP~\cite{traceccsp} and Sagas~\cite{compsursagas,newcomp}. These compensating process calculi are either interaction-based or flow-composition calculi~\cite{cspsaga}. Interaction-based calculi associate with each transaction explicit compensation sequences, and new compensations are installed in the system as an update to the compensation process by using explicit primitives like  ($inst\left\lfloor \right\rfloor$)~\cite{l2011}. In flow-composition calculi the compensation sequence is built as a composition of smaller compensable components. Attached to each subprocess is a compensation component; when this subprocess is successfully terminated its compensation is composed (sequentially or in parallel) to the compensation sequence. This is to undo the effects of this subprocess in case of system failure.

We are interested in flow-composition calculi, where when a transaction fails, compensations are activated in the order specified by the recovery mechanisms. We distinguish between \textit{static recovery}, which is the activation of a previously implemented compensation sequence, and \textit{dynamic recovery}, which is the activation of a dynamically generated compensation sequence (i.e., generated while the system is running). In turn, dynamic recovery mechanisms can be classified as: \textit{parallel recovery}, if all compensations are executed in parallel; \textit{backward recovery}, if parallel processes are compensated in parallel and sequential processes are compensated in backward order; and \textit{general dynamic recovery}, which is backward recovery with the option of replacing or discarding compensations at runtime~\cite{dcpi,l2010,l2011}.

Specifically, we focus on CSP as a modelling language for complex systems,
because of its flexible communication patterns and because it provides
simple reasoning mechanisms to verify significant properties of models,
such as good/bad traces, deadlock freedom and divergence.

Compensation has been introduced in CSP by Butler, Hoare, and Ferreira
who defined compensating CSP (cCSP)
as a flow-composition calculus with a backward recovery
mechanism \cite{traceccsp}. cCSP has been extended  by Chen, Liu, and Wang
in the  Extended compensating
CSP (EcCSP),  bringing back some significant operators from the
original CSP and developing a theory of refinement \cite{newccsp1,newccsp2}.

In this paper we extend cCSP further by introducing primitives to facilitate general dynamic recovery. We call the new calculus DEcCSP (Dynamic EcCSP). Improving the recovery mechanism from backward recovery to general dynamic recovery allows compensations to be replaced or discarded after they have been recorded.  This is useful in many cases, such as: (i)~The compensation process is unknown at the start. (ii)~The compensation process is subject to change while the process evolves. (iii)~The compensation's logic is complex and spans several processes. We demonstrate some of these cases in Section~\ref{case}.

Additionally, DEcCSP extends EcCSP by including all the standard CSP operators, to facilitate the specification of complex systems. DEcCSP provides a conditional (if-then-else), iteration (while-do), prefixing operator, named processes and channels.
Channels have been used informally in extensions of
CSP~\cite{traceccsp,opeccsp,newccsp1,newccsp2,riponthesis}.  We have
extended the syntax of DEcCSP with primitives for channel
communication, and adapted the semantics to allow processes to pass
data (the latter is omitted due to lack of space).

The remainder of this paper is organised as follows: Section~\ref{sec:cCSP} recalls cCSP. Section~\ref{sec:EcCSP} provides an operational semantics for EcCSP, which had so far only a denotational semantics. The operational semantics for EcCSP is used as a basis for the design, in section~\ref{sec:new}, of the syntax and the operational semantics of our calculus, DEcCSP. Section~\ref{case} illustrates its expressive power using a case study. Finally, Section~\ref{sec:conclusions}
concludes the paper and briefly discusses future directions.
%__________________________________________________________________________________________________
\section{Background: compensating CSP (cCSP)}
\label{sec:cCSP}
In this section we recall the main concepts in cCSP~\cite{traceccsp}, assuming the reader is familiar with CSP~\cite{csp2,impcsp}.

The novelty in cCSP is the introduction of transaction processing features within the standard CSP processes. cCSP categorises processes into two types: standard processes, which are a subset of standard CSP processes, and compensable processes, where a standard process is attached to another standard process to undo its effects. When a standard process terminates normally, it evolves to $\emptyset$, which means nothing to do; however, when a compensable process terminates normally then its compensation will be preserved in case the transaction fails and the system needs to roll-back.

The syntax of  cCSP is summarised in Figure~\ref{cs}. We describe below the main constructs.

The operator \textbf{ [ ] } is used to identify transaction's boundaries in cCSP. Transactions can be nested; subtransactions should be aborted if the parent transaction is terminated. According to the semantics of cCSP, transaction blocks cannot be interrupted. cCSP includes operators for handling the key concepts of compensations presented in the introduction. To represent unsuccessful termination of a process, new terminal signals (events) have been added to the calculus: \textbf{($!$)} represents internal fault; \textbf{($?$)} represents yielding to external fault. In addition to these terminal events new primitive processes have been introduced: $\THROW$ is a process that throws an exception then terminates; $\YIELD$ is a process that yields to an external exception and terminates. A new operator  \textbf{$\triangleright$} has been added to implement fault handler (named interrupt handler). In $p \triangleright q$, $q$ is the fault handler of $p$. Compensable process can be defined in the calculus as a pair of standard processes which are composed with the new operator \textbf{$\div$}. In $p \div q$, $q$ is the compensation handler of $p$. The $\emptyset$ primitive process is added to the syntax of cCSP to represent a process which does nothing. It is equal to $\STOP$ in the original CSP.
cCSP processes can be composed sequentially or in parallel.  Parallel processes can synchronise on terminal events solely. Choice between two processes is resolved by either of them performing an event.

\begin{figure}
{\small
\begin{tabular}{llllll}
\multicolumn{3}{c}{(Standard Processes)} &\multicolumn{3}{c}{(Compensable Processes)}  \\
$p,q ::=$ & $a$   & (Atomic process)  & $pp,qq  ::=$  & $p \div q$   & (Compensation pair (CP)) \\
& $| p \square q $ & (Choice operator) & & $|pp \square qq $ & (Choice operator)  \\
& $|p;q$&(Sequential composition)&&$|pp;qq\;$&(Sequential composition)\\
& $|p\parallel q$&(Parallel composition)&&$|pp\parallel qq$&(Parallel composition)\\
& $|\SKIP$&(Primitive process)&&$|\SKIPP$&(Primitive process)\\
& $|\THROW$&(Primitive process)&&$|\THROWW$&(Primitive process)\\
& $|\YIELD$&(Primitive process)&&$|\YIELDD$&(Primitive process)\\
& $|p\triangleright q$&(Interrupt handler)&&\\
& $|\emptyset$&(Primitive process)&&\\
& $|[pp]$&(Transaction block)&&\\
\end{tabular}}
\caption{cCSP Syntax}
\label{cs}
\end{figure}

We are now going to describe the operational semantics of these operators and primitive processes; it is based on the semantics presented by Ripon and Butler~\cite{opeccsp}, but we have adapted it to follow the operational semantics of the standard CSP~\cite{csp2,impcsp}. Throughout what follows, the following notations will be used. Standard processes will be referred to by using lower case letters $p$, $q$. Compensable processes will be referred to by using double letters $pp$, $qq$. The set $\Sigma$ is the universal set that contains all the observable events in a system; $a$, $b$,... will be used to range over this set. The set $\Omega$ consists of the terminal events $\{!,?,\surd \}$; $\omega$ will be used to range over this set. We define $\Sigma^\tau:= \Sigma \cup \{\tau\}$ and $\Sigma^{\tau\Omega} := \Sigma^\tau\cup\Omega$. Finally, capital letters $A$, $B$,.. will denote sets of observable events.

The following are \emph{standard processes}: \emph{primitive processes} are
{\small\AxiomC{\rule{0em}{1em}}
\UnaryInfC{$\SKIP \stackrel{\surd}{\longrightarrow} \emptyset$}
{\DisplayProof}},
{\small\AxiomC{\rule{0em}{1em}}
\UnaryInfC{$\THROW \stackrel{!}{\longrightarrow} \emptyset$}
{\DisplayProof}}, and
{\small\AxiomC{\rule{0em}{1em}}
\UnaryInfC{$\YIELD \stackrel{\omega}{\longrightarrow} \emptyset$}
{\DisplayProof}}
for $\omega \in  \{?,\surd \}$. For $a\in\Sigma$, the following is process $a$:
{\small\AxiomC{\rule{0mm}{1em}}
\UnaryInfC{$a \stackrel{a}{\longrightarrow} \SKIP$}
{\DisplayProof}}. If $a,b \in  \Sigma^{\omega\tau}$, then the following two processes are called \emph{Choice}:
{\small\AxiomC{$p \stackrel{a}{\longrightarrow} p'$}
\UnaryInfC{$p \, \square \, q \stackrel{a}{\longrightarrow} p'$}
{\DisplayProof}}
and
{\small\AxiomC{$q \stackrel{b}{\longrightarrow} q'$}
\UnaryInfC{$p \, \square \, q \stackrel{b}{\longrightarrow} q'$}
{\DisplayProof}}.
If $a \in \Sigma^\tau$ and $\omega \in \{!,?\}$, the following three processes are called
\emph{Sequential Composition}:
{\small\AxiomC{$p \stackrel{a}{\longrightarrow} p'$}
\UnaryInfC{$p \, ; \, q \stackrel{a}{\longrightarrow} p'\, ; \, q$}
{\DisplayProof}},
{\small\AxiomC{$p \stackrel{\surd}{\longrightarrow} p'$}
\UnaryInfC{$p \, ; \, q \stackrel{\tau}{\longrightarrow} q$}
{\DisplayProof}}, and
{\small\AxiomC{$p \stackrel{\omega}{\longrightarrow} \emptyset$}
\UnaryInfC{$p \, ; \, q \stackrel{\omega}{\longrightarrow} \emptyset$}
{\DisplayProof}}.
For $a \in \Sigma^\tau$ and
$\omega \in \{\surd,?\}$, the following three processes are called \emph{Interrupt Handler}:
{\small\AxiomC{$p \stackrel{a}{\longrightarrow} p'$}
\UnaryInfC{$p \, \triangleright \, q \stackrel{a}{\longrightarrow} p'\, \triangleright \, q$}
{\DisplayProof}},
{\small\AxiomC{$p \stackrel{!}{\longrightarrow} p'$}
\UnaryInfC{$p \, \triangleright \, q \stackrel{\tau}{\longrightarrow} q$}
{\DisplayProof}},
and
{\small\AxiomC{$p \stackrel{\omega}{\longrightarrow} \emptyset$}
\UnaryInfC{$p \, \triangleright \, q \stackrel{\omega}{\longrightarrow} \emptyset$}
{\DisplayProof}}.
We define the binary operation $(\omega,\omega')\mapsto \omega\&\omega'$ by the following table:
\begin{center}\begin{tabular}{cccc}
 & $!$ & $?$ & $\surd$\\
 \midrule
$!$ & $!$ & $!$ & $!$\\
$?$ &  $!$ & $?$ & $?$\\
$\surd$ & $!$ & $?$ & $\surd$\\
\end{tabular}\end{center}
Then, for $b,c  \in \, \Sigma^\tau$ and $\omega,\omega' \in \, \Omega$, the following three processes are called \emph{Parallel Composition}:
{\small\AxiomC{$p \stackrel{b}{\longrightarrow} p'$}
\UnaryInfC{$p \, \parallel \, q \stackrel{b}{\longrightarrow} p'\, \parallel \, q$}
{\DisplayProof}},
{\small\AxiomC{$q \stackrel{c}{\longrightarrow} q'$}
\UnaryInfC{$p \, \parallel \, q \stackrel{c}{\longrightarrow} p \, \parallel \, q'$}
{\DisplayProof}},
and
{\small\AxiomC{$p \stackrel{\omega}{\longrightarrow} \emptyset$}
\AxiomC{$q \stackrel{\omega'}{\longrightarrow} \emptyset$}
\BinaryInfC{$p \, \parallel \, q \stackrel{\omega \& \omega'}{\longrightarrow} \emptyset$}
{\DisplayProof}}.
Finally, the following three processes are called \emph{Transaction block}:
{\small\AxiomC{$pp \stackrel{a}{\longrightarrow} pp'$}
\UnaryInfC{$[ pp ] \stackrel{a}{\longrightarrow} [pp']$}
{\DisplayProof}},
{\small\AxiomC{$pp \stackrel{!}{\longrightarrow} p$}
\UnaryInfC{$[ pp ] \stackrel{!}{\longrightarrow} p$}
{\DisplayProof}},
and
{\small\AxiomC{$pp \stackrel{\surd}{\longrightarrow} p$}
\UnaryInfC{$[ pp ] \stackrel{\surd}{\longrightarrow} \emptyset$}
{\DisplayProof}}.
This finishes the list of \emph{standard processes}.

\medskip

We now define the \emph{compensable processes}: For $\omega \in \{!,?\}$, we call the following three processes \emph{compensation pair}:
{\small\AxiomC{$p \stackrel{a}{\longrightarrow} p'$}
\UnaryInfC{$p \, \div \, q \stackrel{a}{\longrightarrow} p'\, \div \, q$}
{\DisplayProof}},
{\small\AxiomC{$p \stackrel{\surd}{\longrightarrow} \emptyset$}
\UnaryInfC{$p \, \div \, q \stackrel{\surd}{\longrightarrow} q$}
{\DisplayProof}},
and
{\small\AxiomC{$p \stackrel{\omega}{\longrightarrow} \emptyset$}
\UnaryInfC{$p \, \div \, q \stackrel{\omega}{\longrightarrow} \SKIP$}
{\DisplayProof}}.
Furthermore, for
$\omega \in \{?,\surd\}$, the following are \emph{primitive processes}: $\SKIPP = \SKIP \div \SKIP$, $\THROWW = \THROW \div \SKIP$, $\YIELDD = \YIELD \div \SKIP$,
{\small\AxiomC{\rule{0em}{1em}}
\UnaryInfC{$\SKIPP \stackrel{\surd}{\longrightarrow} \SKIP$}
{\DisplayProof}},
{\small\AxiomC{\rule{0em}{1em}}
\UnaryInfC{$\THROWW \stackrel{!}{\longrightarrow} \SKIP$}
{\DisplayProof}}, and
{\small\AxiomC{\rule{0em}{1em}}
\UnaryInfC{$\YIELDD \stackrel{\omega}{\longrightarrow} \SKIP$}
{\DisplayProof}}.

If
$\omega,\omega' \in \Omega$ and
$a,b \in \Sigma^\tau$,
we call the following processes \emph{Choice}:
{\small\AxiomC{$pp \stackrel{a}{\longrightarrow} pp'$}
\UnaryInfC{$pp \, \square \, qq \stackrel{a}{\longrightarrow} pp'$}
{\DisplayProof}},\linebreak
{\small\AxiomC{$qq \stackrel{b}{\longrightarrow} qq'$}
\UnaryInfC{$pp \, \square \, qq \stackrel{b}{\longrightarrow} qq'$}
{\DisplayProof}},
{\small\AxiomC{$pp \stackrel{\omega}{\longrightarrow} p$}
\UnaryInfC{$pp \, \square \, qq \stackrel{\omega}{\longrightarrow} p$}
{\DisplayProof}}, and
{\small\AxiomC{$qq \stackrel{\omega'}{\longrightarrow} q$}
\UnaryInfC{$pp \, \square \, qq \stackrel{\omega'}{\longrightarrow} q$}
{\DisplayProof}}.
If $a \in \Sigma^\tau$ and $\omega \in \{!,?\}$, then the following three processes are called
\emph{Sequential Composition}:
{\small\AxiomC{$pp \stackrel{a}{\longrightarrow} pp'$}
\UnaryInfC{$pp \, ; \, qq \stackrel{a}{\longrightarrow} pp'\, ; \, qq$}
{\DisplayProof}},
{\small\AxiomC{$pp \stackrel{\surd}{\longrightarrow} p$}
\UnaryInfC{$pp \, ; \, qq \stackrel{\tau}{\longrightarrow} \langle qq,p\rangle$}
{\DisplayProof}}, and
{\small\AxiomC{$pp \stackrel{\omega}{\longrightarrow} p$}
\UnaryInfC{$pp \, ; \, qq \stackrel{\omega}{\longrightarrow} p$}
{\DisplayProof}}.
For $a \in \Sigma^\tau$ and $\omega \in \Omega$, the following two processes are called
\textit{the auxiliary operator:}
{\small\AxiomC{$qq \stackrel{a}{\longrightarrow} qq'$}
\UnaryInfC{$\langle qq,p\rangle \stackrel{a}{\longrightarrow} \langle qq',p\rangle$}
{\DisplayProof}}
and
{\small\AxiomC{$qq \stackrel{\omega}{\longrightarrow} q$}
\UnaryInfC{$\langle qq,p\rangle \stackrel{\omega}{\longrightarrow} q \, ; \, p$}
{\DisplayProof}}. Finally, if $b,c  \in \, \Sigma^\tau$ and $\omega,\omega' \in \, \Omega$, then the following three processes are called \emph{Parallel Composition}:
{\small\AxiomC{$pp \stackrel{b}{\longrightarrow} pp'$}
\UnaryInfC{$pp \, \parallel \, qq \stackrel{b}{\longrightarrow} pp'\, \parallel \, qq$}
{\DisplayProof}},
{\small\AxiomC{$qq \stackrel{c}{\longrightarrow} qq'$}
\UnaryInfC{$pp \, \parallel \, qq \stackrel{c}{\longrightarrow} pp \, \parallel \, qq'$}
{\DisplayProof}}\nolinebreak, and
{\small\AxiomC{$pp \stackrel{\omega}{\longrightarrow} p \quad qq \stackrel{\omega'}{\longrightarrow} q$}
\UnaryInfC{$pp \, \parallel \, qq \stackrel{\omega \& \omega'}{\longrightarrow} p \, \parallel \, q$}
{\DisplayProof}}.

\section {Extended cCSP (EcCSP)}
\label{sec:EcCSP}
Chen, Liu, and Wang
extended cCSP, adapted its trace semantics and developed stable-failure semantics and failure-divergence semantics for cCSP as in the standard CSP \cite{newccsp1,newccsp2}. They also brought back to the syntax of cCSP the original CSP operators: hiding, renaming, non-deterministic choice and recursion. In addition, they changed the parallel operator to be synchronous, and introduced speculative choice ($\boxtimes$). A preliminary semantics for speculative choice was presented in \cite{traceccsp,riponthesis}, however, it was not included in the original cCSP. The EcCSP syntax is summarised in Figure~\ref{es}.

\begin{figure}
{\small
\begin{tabular}{llllll}
\multicolumn{3}{c}{(Standard Processes)} &\multicolumn{3}{c}{(Compensable Processes)}  \\
$p,q ::=$ & $\ldots$   & (cCSP syntax) & $pp,qq  ::=$  & $\ldots$   & (cCSP syntax) \\
& $| p \sqcap q$ & (Internal choice) & & $|pp \sqcap qq $ & (Internal choice)  \\
& $|p \underset{A}{\parallel} q $&(Parallel composition)&&$|pp \underset{A}{\parallel} qq$&(Parallel composition)\\
& $| p \backslash A$ & (Hiding operator) & & $|pp \backslash A $ & (Hiding operator)  \\
& $| p  \llbracket R \rrbracket $ & (Renaming operator) & & $|pp \llbracket R \rrbracket $ & (Renaming operator)  \\
& $| \mu p.f(p) $ & (Recursion)  & & $|\mu pp.ff(pp) $ & (Recursion)   \\
&&&& $|pp \boxtimes qq $ & (Speculative choice)  \\
\end{tabular}}
\caption{EcCSP Syntax}
\label{es}
\end{figure}

EcCSP was developed using denotational semantics in~\cite{newccsp1,newccsp2} and no operational semantics was provided. Therefore, we developed an operational semantics for EcCSP based on the operational semantics of cCSP and CSP. We below describe the new and the adaptive inference rules; cCSP rules which are part of cCSP operational semantics and not listed here are adopted without modification.

The adaptive operators of \textit{standard processes} are as follows:
Atomic process semantics has been adjusted to permit interruptions before or after performing its event. Therefore, for $a\in\Sigma$, if a completed process $a$ is
{\small\AxiomC{\rule{0mm}{1em}}
\UnaryInfC{$a \stackrel{a}{\longrightarrow} \SKIP$}
{\DisplayProof}}, then an interrupted process $a$ is as follows: \textit{before event $a$:}
{\small\AxiomC{\rule{0mm}{1em}}
\UnaryInfC{$a \stackrel{?}{\longrightarrow} \STOP$}
{\DisplayProof}}, \quad
\textit{after event $a$:}
{\small\AxiomC{\rule{0mm}{1em}}
\UnaryInfC{$a \stackrel{a}{\longrightarrow} \STOP$}
{\DisplayProof}}.

Choice, which can be deterministic or non-deterministic as in the standard CSP. Deterministic choice is resolved by either of the processes performing an observable or terminal event as follows.
If $a,b \in  \Sigma^{\omega}$, then the following four processes are called \emph{ External (Deterministic)  Choice}:
{\small\AxiomC{$p \stackrel{a}{\longrightarrow} p'$}
\UnaryInfC{$p \, \square \, q \stackrel{a}{\longrightarrow} p'$}
{\DisplayProof}},
{\small\AxiomC{$q \stackrel{b}{\longrightarrow} q'$}
\UnaryInfC{$p \, \square \, q \stackrel{b}{\longrightarrow} q'$}
{\DisplayProof}},
{\small\AxiomC{$p \stackrel{\tau}{\longrightarrow} p'$}
\UnaryInfC{$p \, \square \, q \stackrel{a}{\longrightarrow} p' \, \square \, q$}
{\DisplayProof}},
and
{\small\AxiomC{$q \stackrel{\tau}{\longrightarrow} q'$}
\UnaryInfC{$p \, \square \, q \stackrel{b}{\longrightarrow} p \, \square \, q'$}
{\DisplayProof}}.

On the other hand, non-deterministic choice is resolved by either of the processes performing the silent event ``$\tau$'' as follows:
{\small\AxiomC{\rule{0mm}{1em}}
\UnaryInfC{$p \, \sqcap \, q \stackrel{a}{\longrightarrow} p$}
{\DisplayProof}},
and
{\small\AxiomC{\rule{0mm}{1em}}
\UnaryInfC{$p \, \sqcap \, q \stackrel{\tau}{\longrightarrow} q$}
{\DisplayProof}}.

The parallel operator, which has been parameterised with a set of events. The purpose of this set is to govern the synchronisation between participants. Thereby, every event in this set should be performed simultaneously, other events can interleave in any order.
If $b,c  \in \, \Sigma^\tau$, and $\omega,\omega' \in \, \Omega$, then the following processes are defining the new \emph{Parallel Composition}:
For $b,c  \notin \, A$,
{\small\AxiomC{$p \stackrel{b}{\longrightarrow} p' $}
\UnaryInfC{$p \, \underset{A}{\parallel} \, q \stackrel{b}{\longrightarrow} p'\, \underset{A}{\parallel} \, q$}
{\DisplayProof}}
and
{\small\AxiomC{$q \stackrel{c}{\longrightarrow} q'$}
\UnaryInfC{$p \, \underset{A}{\parallel} \, q \stackrel{c}{\longrightarrow} p \, \underset{A}{\parallel} \, q' $}
{\DisplayProof}}.
For $a  \in \, A$,
{\small\AxiomC{$p \stackrel{a}{\longrightarrow} p' $}
\AxiomC{$q \stackrel{a}{\longrightarrow} q' $}
\BinaryInfC{$p \, \underset{A}{\parallel} \, q \stackrel{b}{\longrightarrow} p'\, \underset{A}{\parallel} \, q$}
{\DisplayProof}}.
If the binary operation $(\omega,\omega')\mapsto \omega\&\omega'$ as defined in Section \ref{sec:cCSP}, then
{\small\AxiomC{$p \stackrel{\omega}{\longrightarrow} \STOP$}
\AxiomC{$q \stackrel{\omega'}{\longrightarrow} \STOP$}
\BinaryInfC{$p \, \parallel \, q \stackrel{\omega \& \omega'}{\longrightarrow} \STOP$}
{\DisplayProof}}.
Transaction block semantics has been adjusted to permit interruptions by adding the following rule:
{\small\AxiomC{$pp \stackrel{?}{\longrightarrow} p$}
\UnaryInfC{$[ pp ] \stackrel{?}{\longrightarrow} p$}
{\DisplayProof}}.

The new operators of \textit{standard processes} are as follows:
\emph{Hiding}, where a predefined set of events turns to ``$\tau$'' in the targeted process.
If $b \notin \, (A\subseteq \Sigma)$, then
{\small\AxiomC{$ p \stackrel{b}{\longrightarrow}  p'$}
\UnaryInfC{$ p \, \backslash \, A \stackrel{b}{\longrightarrow} p'\, \backslash \, A$}
{\DisplayProof}},
if $a \in \, (A\subseteq \Sigma)$, then
{\small\AxiomC{$ p \stackrel{a}{\longrightarrow}  p' $}
\UnaryInfC{$ p \, \backslash \, A \stackrel{\tau}{\longrightarrow} p'\, \backslash \, A$}
{\DisplayProof}},
and finally if $\omega \in \Omega$, then
{\small\AxiomC{$  p \stackrel{\omega}{\longrightarrow}  \STOP $}
\UnaryInfC{$ p \, \backslash \, A \stackrel{\omega} {\longrightarrow} \STOP$}
{\DisplayProof}}.
\emph{Renaming},
where the events of a process (or a subset of them) are renamed according to a
renaming relation (in other words, if $ R \subseteq \, \Sigma \, \times \,
\Sigma$ is a defined renaming relation which maps events in set $A$ to the
events in set $B$, then renaming process $p$ with $R$ is mapping its
events from $A$ to $B$). The following processes define the \textit{Renaming operator}.
If $(a \, R \, b)$, then
{\small\AxiomC{$  p \stackrel{a}{\longrightarrow} p' $}
\UnaryInfC{$p \llbracket R \rrbracket \stackrel{b}{\longrightarrow} p' \llbracket R \rrbracket$}
{\DisplayProof}},
{\small\AxiomC{$p \stackrel{\tau}{\longrightarrow}  p' $}
\UnaryInfC{$p \llbracket R \rrbracket \stackrel{\tau}{\longrightarrow} p' \llbracket R \rrbracket$}
{\DisplayProof}},
and if $\omega \in \Omega$, then
{\small\AxiomC{$p \stackrel{\omega}{\longrightarrow}  \STOP $}
\UnaryInfC{$p \llbracket R \rrbracket \stackrel{\omega}{\longrightarrow} \STOP$}
{\DisplayProof}}.
\emph{Recursion}, using a fixed point operator as follows:
{\small\AxiomC{\rule{0mm}{1em}}
\UnaryInfC{$\mu p.f(p)  \stackrel{\tau}{\longrightarrow} f[\mu p.f(p) / p]$}
{\DisplayProof}}.
Symmetrically,  we update the choice and parallel composition operators for compensable processes as we did for standard processes as following:

If $\omega,\omega' \in \Omega$ and $a,b \in \Sigma$,
the following six processes are called \emph{External (Deterministic) Choice}:
{\small\AxiomC{$pp \stackrel{a}{\longrightarrow} pp'$}
\UnaryInfC{$pp \, \square \, qq \stackrel{a}{\longrightarrow} pp'$}
{\DisplayProof}},
{\small\AxiomC{$qq \stackrel{b}{\longrightarrow} qq'$}
\UnaryInfC{$pp \, \square \, qq \stackrel{b}{\longrightarrow} qq'$}
{\DisplayProof}},
{\small\AxiomC{$pp \stackrel{\omega}{\longrightarrow} p$}
\UnaryInfC{$pp \, \square \, qq \stackrel{\omega}{\longrightarrow} p$}
{\DisplayProof}},
{\small\AxiomC{$qq \stackrel{\omega'}{\longrightarrow} q$}
\UnaryInfC{$pp \, \square \, qq \stackrel{\omega'}{\longrightarrow} q$}
{\DisplayProof}},
{\small\AxiomC{$pp \stackrel{\omega}{\longrightarrow} p$}
\UnaryInfC{$pp \, \square \, qq \stackrel{\omega}{\longrightarrow} p$}
{\DisplayProof}},\linebreak
and
{\small\AxiomC{$qq \stackrel{\omega'}{\longrightarrow} q$}
\UnaryInfC{$pp \, \square \, qq \stackrel{\omega'}{\longrightarrow} q$}
{\DisplayProof}};
furthermore, the following two processes are called \emph{Internal (Non-deterministic) Choice}:
{\small\AxiomC{$pp \stackrel{\tau}{\longrightarrow} pp'$}
\UnaryInfC{$ pp \, \square \, qq \stackrel{\tau}{\longrightarrow} pp' \, \square \, qq $}
{\DisplayProof}}
and
{\small\AxiomC{$qq \stackrel{\tau}{\longrightarrow} qq'$}
\UnaryInfC{$ pp \, \square \, qq \stackrel{\tau}{\longrightarrow} pp \, \square \, qq'$}
{\DisplayProof}}.

If $b,c  \in \, \Sigma^\tau$ and $\omega,\omega' \in \, \Omega$, then the following three processes are called \emph{Parallel Composition}.
For $b,c  \notin \, A$,
{\small\AxiomC{$pp \stackrel{b}{\longrightarrow} pp'$}
\UnaryInfC{$pp \, \underset{A}{\parallel} \, qq \stackrel{b}{\longrightarrow} pp'\, \underset{A}{\parallel} \, qq$}
{\DisplayProof}},
{\small\AxiomC{$qq \stackrel{c}{\longrightarrow} qq'$}
\UnaryInfC{$ pp \, \underset{A}{\parallel} \, qq \stackrel{c}{\longrightarrow} pp \, \underset{A}{\parallel} \, qq'$}
{\DisplayProof}};
for $a  \in \, A$,
{\small\AxiomC{$pp \stackrel{a}{\longrightarrow} pp'$}
\AxiomC{$ qq \stackrel{a}{\longrightarrow} qq'$}
\BinaryInfC{$ pp \, \underset{A}{\parallel} \, qq \stackrel{a}{\longrightarrow} pp'\, \underset{A}{\parallel} \, qq'$}
{\DisplayProof}};
and finally, for $\omega,\omega' \in \, \Omega$,
{\small\AxiomC{$pp \stackrel{\omega}{\longrightarrow} p$}
\AxiomC{$qq \stackrel{\omega'}{\longrightarrow} q$}
\BinaryInfC{$pp \, \underset{A}{\parallel}  \, qq \stackrel{\omega \& \omega'}{\longrightarrow} p \, \underset{A}{\parallel}  \, q$}
{\DisplayProof}}.

Additionally, we define a new processes to implement the new operators \textit{(Recursion, Hiding, and Renaming)} for compensable processes as we did for standard processes as following:
\emph{Recursion} is the process
{\small\AxiomC{\rule{0mm}{1em}}
\UnaryInfC{$\mu pp.ff(pp)  \stackrel{\tau}{\longrightarrow} ff[\mu pp.ff(pp) / pp]$}
{\DisplayProof}}.
\emph{Hiding} are the following processes:
If $b \notin \, (A\subseteq \Sigma)$, then
{\small\AxiomC{$pp \stackrel{b}{\longrightarrow}  pp'$}
\UnaryInfC{$pp \, \backslash \, A \stackrel{b}{\longrightarrow} pp'\, \backslash \, A$}
{\DisplayProof}},
and
if $a \in \, (A\subseteq \Sigma)$, then
{\small\AxiomC{$ pp \stackrel{a}{\longrightarrow}  pp' $}
\UnaryInfC{$  pp \, \backslash \, A \stackrel{\tau}{\longrightarrow} pp'\, \backslash \, A$}
{\DisplayProof}},
and finally if $\omega \in \Omega$, then
{\small\AxiomC{$  pp \stackrel{\omega}{\longrightarrow}  p $}
\UnaryInfC{$ pp \, \backslash \, A \stackrel{\omega} {\longrightarrow} p \, \backslash \, A$}
{\DisplayProof}}.
Finally, \emph{Renaming} are the following processes:
If $(a \, R \, b)$, then
{\small\AxiomC{$ pp \stackrel{a}{\longrightarrow} pp'$}
\UnaryInfC{$ pp \llbracket R \rrbracket \stackrel{b}{\longrightarrow} pp' \llbracket R \rrbracket$}
{\DisplayProof}}\nolinebreak,
{\small\AxiomC{$ pp \stackrel{\tau}{\longrightarrow}  pp'  $}
\UnaryInfC{$ pp \llbracket R \rrbracket \stackrel{\tau}{\longrightarrow} pp' \llbracket R \rrbracket$}
{\DisplayProof}},
and if $\omega \in \Omega$, then
{\small\AxiomC{$pp \stackrel{\omega}{\longrightarrow}  p $}
\UnaryInfC{$p \llbracket R \rrbracket \stackrel{\omega}{\longrightarrow} p \llbracket R \rrbracket$}
{\DisplayProof}}.

We also define the following processes to implement the \emph{Speculative choice}. \emph{Speculative choice} can be defined as if two processes run in parallel without
synchronisation, then the choice is resolved when one of them commits, and the
other should immediately compensate. If both of them fail then the whole
choice will fail and the processes should compensate in parallel.

If $a,b  \in \Sigma$ and $\omega \in \{!,?\}$, then the following six processes are called
\emph{Speculative Choice}:\linebreak
{\small\AxiomC{$pp \stackrel{a}{\longrightarrow} pp'$}
\UnaryInfC{$ pp \, \boxtimes \, qq ~~~\stackrel{a}{\longrightarrow}~~~ pp'\, \boxtimes \, qq$}
{\DisplayProof}},
{\small\AxiomC{$qq \stackrel{b}{\longrightarrow} qq'$}
\UnaryInfC{$pp \, \boxtimes \, qq ~~~\stackrel{b}{\longrightarrow}~~~ pp \, \boxtimes \, qq'$}
{\DisplayProof}},
{\small\AxiomC{$ pp \stackrel{\omega}{\longrightarrow} p $}
\AxiomC{$qq \stackrel{\omega'}{\longrightarrow} q$}
\BinaryInfC{$ pp \, \boxtimes \, qq ~~~\stackrel{\omega \& \omega'}{\longrightarrow}~~~ \langle (\omega \& \omega'),(p \, \parallel \, q) \rangle  $}
{\DisplayProof}},\linebreak
{\small\AxiomC{$pp \stackrel{\surd}{\longrightarrow} p$}
\AxiomC{$ qq \stackrel{\omega}{\longrightarrow} q$}
\BinaryInfC{$ pp \, \boxtimes \, qq \stackrel{\surd}{\longrightarrow} \langle q,p \rangle $}
{\DisplayProof}},
{\small\AxiomC{$qq \stackrel{\surd}{\longrightarrow} q $}
\AxiomC{$  pp \stackrel{\omega}{\longrightarrow} p$}
\BinaryInfC{$ pp \, \boxtimes \, qq \stackrel{\surd}{\longrightarrow} \langle p,q \rangle  $}
{\DisplayProof}}, and
{\small\AxiomC{$pp \stackrel{\surd}{\longrightarrow} p $}
\AxiomC{$qq \stackrel{\surd}{\longrightarrow} q$}
\BinaryInfC{$pp \, \boxtimes \, qq \stackrel{\surd}{\longrightarrow} \langle q,p \rangle \, \square \,  \langle p,q \rangle $}
{\DisplayProof}}.
For $a \in \Sigma^\tau$ and $\omega \in \Omega$, the following two processes are called
\textit{the auxiliary operator:}
{\small\AxiomC{$p \stackrel{a}{\longrightarrow} p'$}
\UnaryInfC{$\langle p,q \rangle \stackrel{a}{\longrightarrow} \langle p',q\rangle$}
{\DisplayProof}}
and
{\small\AxiomC{$p \stackrel{\omega}{\longrightarrow} \STOP$}
\UnaryInfC{$\langle p,q\rangle \stackrel{\omega}{\longrightarrow} q $}
{\DisplayProof}}.

\section{Dynamic Extended cCSP (DEcCSP)}
\label{sec:new}

Improving the compensation recovery mechanism from backward recovery to
general dynamic recovery allows compensations not only to be recorded in
the right order dynamically, but to be discarded or replaced dynamically
too. In this section we extend EcCSP to include primitives that facilitate
general dynamic recovery. The main idea is to use a free process variable
instead of the reverse behaviour process in compensation pairs. The
variable will work as a place holder within the recovery sequence, where
the real content can be retrieved later at the start of the execution.
This will give the designer the ability to replace variable values
whenever needed or discard them if they are no longer needed as long as
the compensation has not been activated yet. Compensation can be discarded
by assigning $\SKIP$ to the variable, where the $\SKIP$ process in fact equals
to an empty compensation.

The use of a process variable to update compensations is inspired by the work of Guidi, Lanese, Montesi, and Zavattaro to model fault handling in SOCK~\cite{sock} (a service-oriented process calculus) \cite{csock}. This idea has also been applied to the $\pi$-calculus~\cite{l2010,l2011}. However, these calculi are interaction-based, whereas DEcCSP is a flow-composition calculus. Because of this difference, we have followed a different strategy to update and discard compensations.

In addition to improving the recovery mechanism, we bring back the remainder of the CSP standard operators. These include: conditional (if-then-else) and iteration (while-do) control blocks, prefixing operator, and named processes. Although control blocks can be simulated in CSP using the primitive operators as Hoare shows~\cite{csp}, Hoare also argues in~\cite{csp} that having a reasonably wide range of operators is needed in practice.

We also introduce channels passing data: we extend the syntax with channel communication primitives, and adapt the semantics to allow processes to pass data (due to lack of space we present only the syntax of channels carrying data, and omit the operational semantics).
Butler, Ripon, Chen, Liu, and Wang  do not include data in their calculi \cite{traceccsp,opeccsp,newccsp1,newccsp2}. Channels have been used informally in Ripon's case study~\cite{riponthesis}.

The rest of this section will be devoted to presenting DEcCSP's syntax and operational semantics in sections~\ref{deccspsyntax} and~\ref{deccspos} respectively.

\subsection{DEcCSP syntax} \label{deccspsyntax}

The syntax of DEcCSP is given in Figure~\ref{des}. DEcCSP extends EcCSP's
syntax with operators to facilitate general dynamic recovery, as explained
above. These operators include: (i)~Assignment, to assign values to
process variables. (ii)Variable compensation pair, where a process
variable will take the place of the compensation in the usual compensation
pair.

DEcCSP also includes the standard CSP operators: \textit{if-then-else},
\textit{while-do}, prefixing operator and named processes $N$. Process
names can be used to specify recursive processes. We also extend the EcCSP
syntax with the standard primitives of channel communications in CSP. We
assume the communicated data are integers or ordinary names.

Events in CSP can be classified into \emph{ordinary names} (a single constant name describing an action to be performed); \emph{compound names}, which are built by composing ordinary names with other ordinary names (e.g, $a.b$) or with an integer expression (e.g, $a.4$) to denote data that is passed without indicating the direction; and \emph{channels}, where the composition operator (.) is replaced by one of (!,?) to indicate the direction of communication via the channel: if $a$ is a channel, then $a?\ell$ denotes input on the variable $\ell$, which will be recorded in the local store, and $a!e$ denotes output of the value of the integer expression $e$.

The traditional input/output notations ($?,!$) coincide with the notations for terminal events ($?,!$) introduced by Butler, Hoare, and Ferreira \cite{traceccsp}, however it will always be clear from the context which one is intended.

\begin{figure}
{\footnotesize
\begin{tabular}{llllll}
\multicolumn{3}{c}{(Standard Processes)} &\multicolumn{3}{c}{(Compensable Processes)}  \\
$p,q ::=$ & $\ldots$   & (EcCSP syntax) & $pp,qq  ::=$  & $\ldots$   & (EcCSP syntax) \\
& $|$ If$\, b \,$ Then $\, p \,$ Else $\, q $& (condition block) & & $|$ If$\, b \,$ Then $\, pp \,$ Else $\, qq $ & (condition block) \\
& $|$ while$\, b \,$  do $\, p \, $ &(Iteration block)&&$|$ while$\, b \,$  do $\, pp \, $ &(Iteration block)\\
& $| N $ &(Process name) & & $| NN  $ &(Process name)  \\
& $| a \longrightarrow p $   & (Prefix operator) & & $| p \div X  $   & (Variable CP) \\
& $|  X := p $ & (Variable assignment)  & & &  \\[3mm]
\multicolumn{3}{c}{(Events)} &\multicolumn{3}{c}{}  \\
$a  ::=$ & \multicolumn{2}{l}{$\mathrm{Names} \quad |\quad a?\ell \quad |\quad a!e \quad |\quad a.e$}\\
$Names ::=$ & \multicolumn{2}{l}{$\mathrm{name} \quad | \quad \mathrm{name}.\mathrm{Names}$}
\end{tabular}}
\caption{DEcCSP Syntax. Here, $e$ is a standard integer expression, $b$ is a standard Boolean expression, and $\ell$ is an integer variable.}
\label{des}
\end{figure}

%__________________________________________________________________________________________________

\subsection{Operational semantics of DEcCSP} \label{deccspos}

We present below the operational semantics of DEcCSP, based on the operational semantics that we developed for EcCSP in Section~\ref{sec:EcCSP}. To deal with variables, we introduce stores to keep track of the different values of these variables. Stores are denoted by $S$, where $S$ is a collection of typed locations. Variables will have a location in this store to hold its current value. The store $S$ is represented as a function $S[\ell \mapsto \nu]$ which associates to each $\ell$ a value $\nu$. We denote the set of locations where $S$ is defined by $dom(S)$.

In our semantics, configurations contain two stores. The first one is a collection of integer locations, and is called the \emph{local} store. We use $\sigma$, $\sigma'$,... to represent its different states. The second one is a collection of process locations, and is called the \emph{global} store. We use $\rho$, $\rho'$,... to represent its different states. We write $\rho(X)$ to denote the value of the process variable $X$ in $\rho$. Configurations are written $((p,\sigma),\rho)$, or $((pp,\sigma),\rho)$.

The local store keeps track of the values of data variables in the scope of the associated process. The global store keeps track of the values of process variables in the full space of configurations. Therefore, the state of the global store is only changed when a new process variable has been declared or if a process variable is assigned a new value.

Below we present the semantics of the new extensions in DEcCSP, the rest of DEcCSP is similar to EcCSP.

\textbf{\textit{General Dynamic recovery}} can be implemented in the calculus by using a \emph{compensation pair with process variable}. A compensation pair with process variable consists of a standard process as a forward behaviour and a process variable as its compensation partner. The variable works as a place holder within the recovery sequence, where the real content can be retrieved later.

A compensation pair with process variable
will be denoted by $px$, to distinguish it from the standard one denoted by $pp$, that is, $px \, = \, p \, \div \, X $, where $p$ is a standard process representing $px$'s forward behaviour, and $X$ is a process variable which works as place holder for $px$'s compensation behaviour. The variable $X$ should be fresh, i.e., not in the domain of the global store.

During the execution of the forward behaviour $p$, $X$'s value can be changed anywhere in the system. If $p$ terminates normally then the variable $X$ will be recorded, and $X$ still can be changed anywhere in the system as long as the compensation sequence has not been activated. If $p$ terminates abnormally then so does the compensation pair, resulting in an empty compensation. This is formalised by the following rules:
{\small\AxiomC{$((p,\sigma),\rho) \stackrel{a}{\longrightarrow} ((p',\sigma'),\rho)$}
\UnaryInfC{$((p \, \div \, X,\sigma),\rho) \stackrel{a}{\longrightarrow} ((p' \, \div \, X,\sigma'),\rho)$}
{\DisplayProof}},
{\small\AxiomC{$((p,\sigma),\rho) \stackrel{\surd}{\longrightarrow} ((\STOP,\sigma),\rho)$}
\UnaryInfC{$((p \, \div \, X,\sigma),\rho) \stackrel{\surd}{\longrightarrow} ((X,\sigma),\rho)$}
{\DisplayProof}}, and
{\small\AxiomC{$ ((p,\sigma),\rho) \stackrel{\omega}{\longrightarrow} ((\STOP,\sigma),\rho)$}
\UnaryInfC{$((p \, \div \, q,\sigma),\rho) \stackrel{\omega}{\longrightarrow} ((\SKIP,\sigma),\rho)$}
{\DisplayProof}}
for $\omega \in \, \{!,?\}$. The value of $X$ can be replaced by assigning a new value to it; to discard the compensation, we assign $\SKIP$:
{\small\AxiomC{\rule{0mm}{1em}}
\UnaryInfC{$((X:=p,\sigma),\rho)\stackrel{\tau}{\longrightarrow} ((\SKIP,\sigma),\rho[X \mapsto p])$}
{\DisplayProof}}\nolinebreak.

The stored value of the process variable $X$ will be retrieved if the associated transaction throws an exception. If a transaction $[ ((pp,\sigma),\rho) ]$ throws an exception $!$, then the transaction block will be ended, and the corresponding compensation $p$ will be activated. Therefore, the values of every process variable should be retrieved by replacing it with its value in the global store. If $ \rho(X)=p$ then
{\small\AxiomC{\rule{0mm}{1em}}
\UnaryInfC{$((X,\sigma),\rho) \stackrel{\tau}{\longrightarrow} ((p,\sigma),\rho)$}
{\DisplayProof}}

\textbf{\textit{Control Blocks.}}  \textit{If-then-else} and \textit{while-do} are the same as the standard control blocks, where $b$ in the two control blocks is a Boolean expression which is evaluated according to the standard Boolean semantics.
The following three processes are defining the \textit{Condition Block} for standard processes:\linebreak
{\small\AxiomC{$((b,\sigma),\rho) \stackrel{\tau}{\longrightarrow} ((b',\sigma'),\rho)$}
\UnaryInfC{$((\mathrm{If}~b~\mathrm{Then}~p~\mathrm{Else}~q,\sigma),\rho)  \stackrel{\tau}{\longrightarrow} ((\mathrm{If}~b'~\mathrm{Then}~p~\mathrm{Else}~q,\sigma'),\rho)$}
{\DisplayProof}},
{\small\AxiomC{\rule{0mm}{1em}}
\UnaryInfC{$((\mathrm{If}~\mathrm{True}~\mathrm{Then}~p~\mathrm{Else}~q,\sigma),\rho)  \stackrel{\tau}{\longrightarrow} ((p,\sigma),\rho)$}
{\DisplayProof}}\nolinebreak,\linebreak
and
{\small\AxiomC{\rule{0mm}{1em}}
\UnaryInfC{$((\mathrm{If}~\mathrm{False}~\mathrm{Then}~p~\mathrm{Else}~q,\sigma),\rho)  \stackrel{\tau}{\longrightarrow} ((q,\sigma),\rho)$}
{\DisplayProof}}.
The following three processes are defining the \textit{Condition Block} for compensable processes:
{\small\AxiomC{$((b,\sigma),\rho) \stackrel{\tau}{\longrightarrow} ((b',\sigma'),\rho)$}
\UnaryInfC{$((\mathrm{If}~b~\mathrm{Then}~pp~\mathrm{Else}~~qq,\sigma),\rho)  \stackrel{\tau}{\longrightarrow} ((\mathrm{If}~b'~\mathrm{Then}~pp~\mathrm{Else}~qq,\sigma'),\rho) $}
{\DisplayProof}},
{\small\AxiomC{\rule{0mm}{1em}}
\UnaryInfC{$((\mathrm{If}~\mathrm{True}~\mathrm{Then}~pp~\mathrm{Else}~qq,\sigma),\rho)  \stackrel{\tau}{\longrightarrow} ((pp,\sigma),\rho)$}
{\DisplayProof}},
and
{\small\AxiomC{\rule{0mm}{1em}}
\UnaryInfC{$((\mathrm{If}~\mathrm{False}~\mathrm{Then}~pp~\mathrm{Else}~qq,\sigma),\rho)  \stackrel{\tau}{\longrightarrow} ((qq,\sigma),\rho)$}
{\DisplayProof}}.

\smallskip

Finally, \textit{Iteration Block} in standard and compensable processes can be defined as follows:
\begin{center}
{\small\AxiomC{\rule{0mm}{1em}}
\UnaryInfC{$((\mathrm{While}~b~\mathrm{Do}~p,\sigma),\rho) \stackrel{\tau}{\longrightarrow} ((\mathrm{If}~b~\mathrm{Then}~(p;\mathrm{While}~b~\mathrm{Do}~p)~\mathrm{Else}~\SKIP,\sigma),\rho)$}
{\DisplayProof}} and\\
{\small\AxiomC{\rule{0mm}{1em}}
\UnaryInfC{$((\mathrm{While}~b~\mathrm{Do}~pp,\sigma),\rho) \stackrel{\tau}{\longrightarrow} ((\mathrm{If}~b~\mathrm{Then}~(pp;\mathrm{While}~b~\mathrm{Do}~pp)~\mathrm{Else}~\SKIP,\sigma),\rho)$}
{\DisplayProof}}.\end{center}

\paragraph{\textit{Named Processes for Definitions and Recursion}.}
We write $(N=p)$ if $N$ is the name of the standard process $p$, and $(NN=pp)$ if $NN$ is the name of the compensable process $pp$. Process names can be used in the more common style of recursion where the process name is used in the process body. This can be defined as following:
If $N=p$ then
{\small{
{\small\AxiomC{\rule{0mm}{1em}}
\UnaryInfC{$((N,\sigma),\rho) \stackrel{\tau}{\longrightarrow} ((p,\sigma),\rho)$}
{\DisplayProof}},}}
and if $NN=pp$ then
{\small{
{\small\AxiomC{\rule{0mm}{1em}}
\UnaryInfC{$((NN,\sigma),\rho) \stackrel{\tau}{\longrightarrow} ((pp,\sigma),\rho)$}
{\DisplayProof}}.}}

\paragraph{\textit{Prefixing}.}
Let $a$ be an event $\in \Sigma \cap \alpha p $ ($\alpha P$ is the set of events that the process can perform), and let $p$ be a process. Prefixing represents a standard process which is ready to engage in event $a$ and then behave as $p$,
{\small\AxiomC{\rule{0mm}{1em}}
\UnaryInfC{$(((a \longrightarrow p, \sigma),\rho) \stackrel{a}{\longrightarrow} ((p,\sigma),\rho))$}
{\DisplayProof}}.

The prefix operator can be used to link critical events, which should be executed in sequence without interruption.
Prefixing differs from sequencing, as the following example shows.

According to DEcCSP's syntax ($P = a \longrightarrow b \longrightarrow \SKIP$) and ($Q = a ; b ; \SKIP$) are valid processes. However, $Q$ can be interrupted  whereas $P$ can not. To be more clear the two process will be composed in parallel with $\THROW$ as follows: $ P \, \underset{\phi}{\|} \, Q \, \underset{\phi}{\|} \, \THROW$

If $\THROW$ has been performed before $P$ or $Q$, then the two processes can be interrupted. If $\THROW$ has happened after, e.g., the first event ($a$), then $Q$ will be prohibited from continuing execution while $P$ will not. This is due to the successful terminal event after $a$ in $Q$. This terminal event will synchronise with the terminal event $!$ in $\THROW$ resulting in $!$ which terminates the parallel operator. Such terminal event does not exist in $P$ therefore the process will continue its execution.

\section{Case study} \label{case}

To illustrate the new features of DEcCSP, in this section  we develop a case study based on the one described in~\cite{opeccsp}. We first demonstrate the basic extensions to cCSP, namely channels passing integers and control blocks. Then, the case study is extended to emphasise the general dynamic recovery.

Customers can order products from a warehouse. A customer should provide an item number, the quantity and his membership number (0 will be sent if the customer is not a member). If the order is accepted then proceed to fulfill this order and subtract the quantity from the inventory. If this action completed but later the transaction fails then the deducted quantity should be returned to the inventory. The {\small {\fontfamily{pcr}\selectfont FulfillOrder}} process starts by booking a courier which should be compensated if the transaction fails by cancelling the courier.  If the customer is a member of this warehouse then no fee is charged, otherwise fees should be paid as  part of the compensation process for booking the courier. {\small {\fontfamily{pcr}\selectfont BookCourier}} runs in parallel with packing the order, where each item ordered is packed, if there is an error then the item should be unpacked. At the same time, the customer's credit card is charged with the amount needed. This is done at the same time because it usually succeeds. However, if the credit process fails then the whole transaction fails and the system should be compensated.

This ordering system runs in parallel with the {\small {\fontfamily{pcr}\selectfont Customer}} process which issues an order or applies for a membership, and a {\small {\fontfamily{pcr}\selectfont Bank}} process which validates the credit card. In the following, we use $\|$ with no parameters to mean that the participant processes synchronise on the terminal events solely, 
and $\parallel_{i=0}^{i=y}$ is an indexed version of the parallel operator between several processes.

\smallskip

\begin{center}
\parbox{.9\textwidth}{
{\small {\fontfamily{pcr}\selectfont
System = ((OrderTransaction  $ \underset{\mathtt{B}}{\|}$   Bank ) $ \underset{\mathtt{A}}{\|}$ (Customer $ \underset{\mathtt{C}}{\|}$ CustomerService))  \newline
Where A=\{Order.x.y.ns \}, B=\{CreditCheck.N, Ok, NotOk\} and \newline C=\{RequestMembership, MembershipNumber.ns \} \newline
OrderTransaction = [ProcessOrder]  \newline
ProcessOrder = ((Order?x?y?ns; deduct.x.y) $\div$ Restock.x.y) ; FulfillOrder \newline
FulfillOrder = BookCourier $\div$ (If  ns=0  Then  cancelcourier1  Else \\  cancelcourier) $\parallel$ ($\parallel_{\mathtt{i}=\mathtt{0}}^{\mathtt{i}=\mathtt{y}}$ (Pack.x $\div$ Unpack.x)) $\parallel$ ((CreditCheck!N ; \\ (Ok $\square$ (NotOk ; THROW)))$\div$ SKIP) \newline
cancelcourier1 = cancelcourier ; penalty \newline
Customer= (Order!x!y!ns ; Customer) $\sqcap$ ((RequestMembership; payfee; \newline MembershipNumber?ns) ; Customer)  \newline
CustomerService=RequestMembership;createprofile;MembershipNumber!ns; \newline
CustomerService \newline
Bank =  CreditCheck?N ; (Ok ; Bank) $\sqcap$ (NotOk ; Bank) \newline
}}
}\end{center}

\smallskip

The system will be started by the customer process either executing {\small {\fontfamily{pcr}\selectfont Order!x!y!ns}}, where $x$, $y$ and $ns$ are integers, or {\small {\fontfamily{pcr}\selectfont RequestMembership}}. {\small {\fontfamily{pcr}\selectfont Order!x!y!ns}} starts a new transaction to process the order, that is, deduct the quantity from the inventory then proceed to {\small {\fontfamily{pcr}\selectfont FulfillOrder}}, which consists of three parallel subprocesses. Due to the parallel operator in this process its execution may have different possibilities. Assume the following scenario: a courier has been booked then three items of the product have been packed before the system checks the credit card. While the system is waiting for the bank to return the answer, a fourth item has been packed.

This scenario will lead us to the choice: {\small {\fontfamily{pcr}\selectfont (Ok $\square$ (NotOk ; THROW))}}. If the bank answered {\small {\fontfamily{pcr}\selectfont Ok}} then the transaction continues, however, if the bank answered {\small {\fontfamily{pcr}\selectfont NotOk}} the current transaction terminates abnormally causing the associated compensations to start.
The current compensations are:

\smallskip

\begin{center}
\parbox{.9\textwidth}{
{ \small {\fontfamily{pcr}\selectfont Unpack.x $\|$ Unpack.x $\|$ Unpack.x $\|$ Unpack.x $\|$ (If  ns=0  Then  cancelcourier1 Else  cancelcourier); restock.x.y}}.
}\end{center}

\smallskip

Alternatively, applying for a membership will be started by {\small {\fontfamily{pcr}\selectfont RequestMembership}} which is followed by a series of actions to process the application and is finished by the event {\small {\fontfamily{pcr}\selectfont MembershipNumber.ns}} which will send the membership number for the customer who issued {\small {\fontfamily{pcr}\selectfont RequestMembership}}.

The If statement provides the designer with the ability to delay resolving the choice of which compensation to start until compensation evaluation. However, the If statement is not sufficient for all cases, e.g, consider the following case.

This system assumes that items are always available in the warehouse. If we extend this system by removing this assumption, then items may not be available in the inventory. In this case, the warehouse should order the unavailable items from its two branches starting by the first one because it is nearer. If both branches fail to satisfy the order then the whole transaction fails.

To design this we add to the system  a {\small{\fontfamily{pcr}\selectfont PrepareOrder}} process which checks if the item is available or not, if not then it orders the item from the two branches in sequence.

We now replace the compensation in {\small{\fontfamily{pcr}\selectfont ProcessOrder}} with a variable $X$, because the compensation is not known at the start; it depends on the {\small{\fontfamily{pcr}\selectfont PrepareOrder}} process.

\smallskip

\begin{center}
\parbox{.9\textwidth}{
{\small {\fontfamily{pcr}\selectfont
\noindent OrderTransaction = [ProcessOrder $\underset{\mathtt{D}}{\|}$ (PrepareOrder $\underset{\mathtt{F}}{\|}$ (Branch1 $\underset{\emptyset}{\|}$ Branch2))]\\
Where \\ D=\{Inventory.x.y, InvOK\} and F=\{okbranch1,okbranch2,nobranch1,nobranch2\}\newline
ProcessOrder = ((Order?x?y?ns; X:= SKIP; Inventory!x!y; InvOK; deduct.x.y)  $\div$  X); FulfillOrder \newline
PrepareOrder= (Inventory?x?y; ((Available; X:=restock.x.y; InvOK) $\square$ \\ (NotAvailable;  branch1!x!y; ((okbranch1;X:=(restock.x.y;Cancelbranch1); \\ InvOK) $\square$ (nobranch1;  branch2!x!y; ((okbranch2; X:=(restock.x.y; \\ Cancelbranch2);InvOK) $\square$ (nobranch2; THROW))))) $\div$ SKIP  \newline
Branch1= (branch1?x?y; (okbranch1 $\sqcap$ nobranch1); Branch1) $\div$ SKIP    \newline
Branch2= (branch2?x?y; (okbranch2 $\sqcap$ nobranch2); Branch2) $\div$ SKIP
}}
}\end{center}

\smallskip

The compensation sequence of the above process is:

\smallskip

\begin{center}
\parbox{.9\textwidth}{
{ \small {\fontfamily{pcr}\selectfont Unpack.x $\|$ Unpack.x $\|$ Unpack.x $\|$ Unpack.x $\|$ (If  ns=0  Then  cancelcourier1 \\ Else  cancelcourier); X}}.
}\end{center}

\smallskip

At the start, $X$ is initialised with $\SKIP$. The event {\small{\fontfamily{pcr}\selectfont InvOK}} is used to ensure that the deduction will not happen unless the order has been fulfilled.

\section{Conclusions and future work}
\label{sec:conclusions}
General dynamic recovery allows compensations to be replaced or discarded in the compensation sequence. This is useful in cases where compensations are not known from the beginning or if they are subject to change while the system is running. DEcCSP is a compensating calculus developed as an
extension of EcCSP, improving the recovery mechanism (from backward recovery to general dynamic recovery) and including all of the CSP standard operators.

We have developed an operational semantics for DEcCSP. A denotational
semantics for the three behavioural models of the standard CSP, as well as
a theory of refinement, is left for future work. We also plan to extend
the functionality of DEcCSP by introducing mobility.

%-----------------------------------------------------------------------DOCUMENT-END
\nocite{*}
\bibliographystyle{eptcs}
\bibliography{AlHumaimeedyFernandez}
\end{document}